\documentclass[conference]{IEEEtran}
\usepackage{amsfonts}
\IEEEoverridecommandlockouts
\usepackage{epsfig}
\usepackage{graphicx}
\usepackage{psfig}
\usepackage{subfigure}
\usepackage{epsf}
\usepackage{epstopdf}

\usepackage[cmex10]{amsmath}
\usepackage{booktabs}
\usepackage{fancyhdr}

\usepackage{algorithm}
\usepackage{algorithmic}
\usepackage{epstopdf,cite,color}
\usepackage{amsthm}
\usepackage{amssymb}
\newcommand{\bc}{\begin{center}}
	\newcommand{\ec}{\end{center}}
\newcommand{\be}{\begin{equation}}
\newcommand{\ee}{\end{equation}}
\newcommand{\bea}{\begin{eqnarray}}
\newcommand{\eea}{\end{eqnarray}}

\renewcommand\IEEEkeywordsname{Index Terms}
\usepackage[letterpaper, left=0.625in, right=0.625in, bottom=1in, top=0.70in]{geometry}
\begin{document}
	\title{Mobility-Aware Offloading and Resource Allocation in MEC-Enabled IoT Networks}
	\author{Han~Hu$^*$\textsuperscript{\S}, Weiwei~Song$^*$\textsuperscript{\S}, Qun~Wang\textsuperscript{\dag}, Fuhui~Zhou\textsuperscript{\ddag}, Rose~Qingyang~Hu\textsuperscript{\dag}\\ 
		$^*$Jiangsu Key Laboratory of Wireless Communications, \\ Nanjing University of Posts and Telecommunications, Nanjing, China, 210000\\
		\textsuperscript{\S}Jiangsu Key Lab of Broadband Wireless Communication and Internet of Things, \\Nanjing University of Posts and Telecommunications, Nanjing, China, 210000\\
		\textsuperscript{\dag}Department of Electrical and Computer Engineering, Utah State University, Logan, UT, USA\\
		\textsuperscript{\ddag}College of Electronic and Information Engineering,\\ Nanjing University of Aeronautics and Astronautics, Nanjing, 210000\\
		Emails: \{han\_h, 1018010111\}@njupt.edu.cn, \{claudqunwang, zhoufuhui\}@ieee.org, rose.hu@usu.edu
	}
	\maketitle
	
	\IEEEpeerreviewmaketitle
\begin{abstract}
Mobile edge computing (MEC)-enabled Internet of Things (IoT) networks have been deemed a promising paradigm to support massive energy-constrained and computation-limited IoT devices. IoT with mobility has found tremendous new services in the 5G era and the forthcoming 6G eras such as autonomous driving and vehicular communications. However, mobility of IoT devices has not been studied in the sufficient level in the existing works. In this paper, the offloading decision and resource allocation problem is studied with mobility consideration. The long-term average sum service cost of all the mobile IoT devices (MIDs) is minimized by jointly optimizing the CPU-cycle frequencies, the transmit power, and the user association vector of MIDs. An online mobility-aware offloading and resource allocation (OMORA) algorithm is proposed based on Lyapunov optimization and Semi-Definite Programming (SDP). Simulation results demonstrate that our proposed scheme can balance the system service cost and the delay performance, and outperforms other offloading benchmark methods in terms of the system service cost.
\end{abstract}
\begin{IEEEkeywords}
edge computing, mobility, Lyapunov optimization, offloading, resource allocation. \end{IEEEkeywords}
\IEEEpeerreviewmaketitle
\section{Introduction}
With the fast and pervasive development of Internet of Things (IoT), we expect massive IoT devices that need to be connected to wireless networks. It is predicted that the global mobile data traffic will increase sevenfold in the next five years, while the number of the global mobile devices will be $12.3$ billion by 2022 \cite{cisco}. Such rapidly growing demands necessitate the development of a new wireless architecture that can provide ubiquitous connectivity to massive mobile IoT devices (MIDs). To that end, small cell networks have become a key technology to support massive connectivity and data capacity \cite{udn1}. Due to the spatial proximity between small BSs and MIDs, this architecture can provide MIDs with better communication qualities, i.e., less energy consumption, better coverage, and higher capacity, especially at the edge of the network \cite{udn3}. 5G infrastructure has facilitated the evolution of the traditional IoT towards the new generation IoT with much higher capabilities to carry new services these days \cite{ultraiot}.  

A fundamental challenge in IoT networks is how to tackle the contention between the resource-hungry applications and resource-restricted MIDs. Mobile edge computing (MEC) has become a promising paradigm to address these issues \cite{tan1}\cite{han1}. By deploying edge servers with high computational and communication capacities closer to the end users, MIDs can offload their computation tasks to the nearby MEC servers so that delay sensitive yet computation intensive services can be supported and energy can be saved for battery driven MIDs. Computation offloading in MEC systems has attracted significant research attention from both academia and industry in recent years \cite{tan2},\cite{fuhui}. Mao \textit{et.al.} \cite{singlemao} proposed an optimal binary offloading algorithm by joint optimizing communication and computational resource allocation. Deng \textit{et.al.} \cite{lyadeng} proposed a dynamic parallel computing algorithm to minimize the response time and packet loss under the limitation of energy queue stability for the green MEC framework. Wang \textit{et.al.} \cite{singlewang} incorporated interference management into binary offloading as well as the allocations of physical resource blocks  and computation resources. However, all the models mentioned above only focus on MEC systems with a single edge node. These architectures are relatively simple and not generally applicable to IoT networks.

There are some unique challenges for computation offloading in a multi-MEC enabled IoT network. First, each MID can be covered by multiple MEC servers and each MID needs to first determine which MEC server to be associated with. User association is very important for offloading as it directly affects communication capacity and computation latency. Different from user association polices in the conventional heterogeneous networks \cite{asso2}, both the communication and computation capacity need to be considered in an MEC-enabled network. Second, due to mobility, an MID may need to re-associate to a different MEC server for offloading from time to time \cite{ultraiot}. The service migration from one MEC to another MEC brings additional operation costs, which needs to be considered when designing an offloading scheme. Thirdly, due to user movement, the future information on channel conditions, user location, and task arrival can be difficult to predict. Thus, the task offloading decision has to be made by considering all these uncertainties. Most existing works about offloading schemes in IoT networks \cite{ultraiotoff1},\cite{ultraiotoff2} have focused on a quasi-static scenario and no service migration cost due to mobility was taken into account, which calls for in-depth study on the computation offloading design for mobile IoT networks. 

Motivated by the above-mentioned challenges, in this paper, we investigate the problem of task offloading and resource allocation in a multi-MEC-enabled mobile IoT network, where computation tasks arrive at the MIDs in a stochastic manner. User association and re-association due to mobility are considered during the task offloading design, and service migration cost is also taken into account. The objective is to minimize the average sum long-term service cost of all the MIDs under the constraints of resource availability, minimum rate requirement, and task queue stability. To solve this problem, we design an online mobility-aware offloading and resource allocation algorithm (OMORA) based on the Lyapunov optimization method and Semi-Definite Programming (SDP), which jointly optimize the transmit power, the CPU-cycle frequencies of MIDs, as well as the user association vector for offloading. Simulation results demonstrate that the proposed algorithm can balance the service cost (the weighted sum of the power consumption and the service migration cost) and the delay performance in the IoT network, and outperforms other offloading benchmark methods in terms of the system service cost.

The rest of the paper is organized as follows. In Section II, the system model is presented. Section III formulates the average service cost minimization problem. In Section IV, an online mobility-aware offloading and resource allocation algorithm is developed. Simulation results are given in Section V. Finally, the paper is concluded in Section VI.

\section{System Model}
As illustrated in Fig. 1, an MEC-enabled IoT network is considered with $M$ densely deployed Small Base Stations (SBS), denoted as $\mathcal{M} = \{ 1,2,...,M\}$, to serve a set of $U$ MIDs, denoted as $\mathcal{U} = \{ 1,2,...,U\}$. Each SBS is equipped with an MEC server to provide computation offloading service to the resource-constrained MIDs, such as smartphones, tablets, and wearable devices. Due to the constrained computation capabilities, each MID can offload partial computation tasks to an MEC server from one of the nearby SBSs it is associate to. 

\label{systemmodel}
\begin{figure}[h]
	\centering
		\includegraphics[width=3.7in]{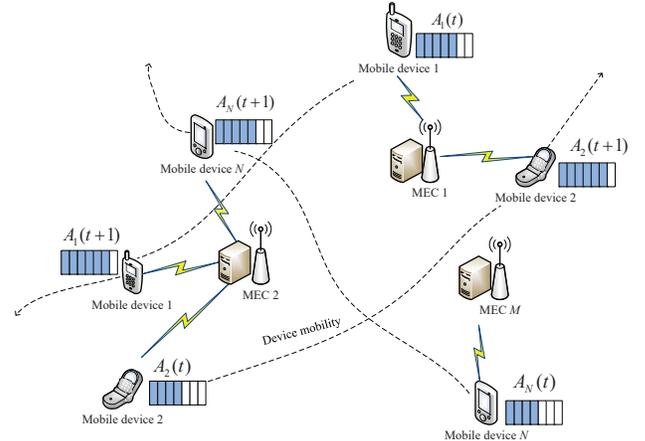}
		\caption{System model.}\label{sym}
\end{figure}

We focus on a multi-user mobility scenario. The MIDs are randomly distributed and move continuously in the network, whereas the MEC servers are static. The system is assumed to operate in a time-slotted structure and the time slot length is $\tau$, i.e. $t \in {\rm \mathcal{T}} = \{ 1,2,...,T\}$. Let the binary indicator $x_u^m(t)$ denote the different user association variable, where $x_u^m(t)=1$ if MID $u \in \mathcal{U}$ chooses to associate with MEC server $m$, otherwise, $x_u^m(t)=0$. Each MID can only associate with one MEC server at one time. The constraints for the user association policy are given as follows:
\be
\sum_{m=1}^M {x_u^m(t) = 1,\forall u \in U,t \in \mathcal{T}}, 
\ee

\be
x_u^m(t) \in \{ 0,1\} ,\forall m \in \mathcal{M},u \in U,t  \in \mathcal{T}.
\ee

The number of MIDs concurrently served by the MEC server $m$ at $t$ is given by $N_m(t)=\sum_{u=1}^U {x_u^m(t)}$, which satisfies
\be
N_m(t) \le N_m^{\max },\forall~m \in \mathcal{M},t \in \mathcal{T}.
\ee

\subsection{Computation Task Queueing Models}
For MID $u \in \mathcal{U}$, let $A_u(t)$ represent the number of the arrival computation tasks. Note that the prior statistical information about $A_u(t)$ is not required to be known, which is usually difficult to be obtained in practical systems. 

At the beginning of each time slot, MID $u$ firstly associates with an appropriate MEC server and executes partial computation tasks $D_u^l(t)$ at the local CPU. Meanwhile $D_u^o(t)$ is offloaded to the associated MEC server. The arrived but not yet processed data is queued in the task buffer for the next or future time slot processing. Let $Q_u(t)$ be the queue backlog of MID $u$ at time slot $t$, and its evolution equation can be expressed as
\be
{Q}_u(t + 1) = \max \{ {Q}_u(t) - {D_u(t)},0\} + {A}_u(t),
\ee
where $D_u(t)=D_u^{o}(t)+D_u^{l}(t)$ is the total executed amount of computation tasks for MID $u$ at time slot $t$.
\subsection{Local Execution Model}
Let $f_u(t)$ denote the local CPU-cycle frequency of MID $u$ with a maximum value $f_{max}$. Let $\gamma_u$ be the computation intensity (in CPU cycles per bit). Thus, the local task processing rate can be expressed as $r_u^l(t)= f_u(t)/\gamma_u$,
The available amount of computation tasks executed locally by MID $u$ at time slot $t$ is  $D_u^l(t)=r_u^l(t)\tau$.

We use the widely adopted power consumption model $P_u^l(t)=\kappa_{mob}f_u(t)^3$ to calculate the power consumption of MID $u$ for local execution, where $\kappa_{mob}$ is the energy coefficient depending on the chip architecture \cite{qunwang}.

\subsection{Task Offloading Model}
The amount of $D_u^o(t)$ at time slot $t$ is offloaded from MID $u$ to its associated MEC through the uplink channel. The wireless channel is assumed to be independent and identically distributed (i.i.d) frequency-flat block fading, i.e., the channel remains static within each time slot, but varies among different time slots. The small-scale Rayleigh fading coefficient from MID $u$ to its serving MEC $m$ is denoted as ${h_{u,m}}(t)$, which is assumed to be exponentially distributed with a unit mean. Thus, the channel power gain from MID $u$ to its serving MEC $m$ can be represented by ${H_{u}^{m}}(t) = {h_{u,m}}(t){g_0}({d_0}/{d_{u,m}})^\theta$, where $g_0$ is the path-loss constant, $\theta$ is the path-loss exponent, $d_0$ is the reference distance, and ${d_{u,m}}$ is the distance from MID $u$ and MEC server $m$. The system uses Frequency Division Multiple Access (FDMA) in each cell and there is no intra-cell interference. According to the Shannon-Hartley formula, the achievable rate of MID $u$ to its associated MEC server at time slot ${t}$ is given as
  \be
 r_{u}^o(t)=\sum_{m=1}^M {x_u^m(t)}\omega{\log_2}(1+\frac{{{H_{u}^{m}}(t)p_u^{tx}(t)}}{\chi+ \sigma ^2}), 
 \ee
 where $\omega$ is the system bandwidth of each MID. ${\sigma ^{\rm{2}}}$ is the background noise variance and the variable $\chi$ is the average inter-cell interference power which is assumed to be a constant by applying intelligent interference management scheme according to the different cell size \cite{inter2}\cite{inter1}. Then, the available amount of computation tasks offloaded from MID $u$ to its associated MEC is $D_u^o(t)=r_u^o(t)\tau$.
 The power consumption for offloading is 
 \be	
 P_u^o(t) = \zeta p_u^{tx}(t)+p_r,
 \ee
where $\zeta$ is the amplifier coefficient and $p_r$ is the constant circuit power consumption.

\subsection{Service Migration Cost Model}
With user mobility, the associated MEC server changes from time to time in order to best serve the user. However, the handover results in an additional cost. Especially, when transferring the service profile of each MID across edges, it incurs extensive usage of the network resources and also increase the energy consumption of network devices \cite{ouyang}.
To model the service migration cost incurred by the handover, let ${c_u}(t)$ be the service migration cost from source MEC server $j \in {\mathcal{M}}$ at $t-1$ to the target MEC server $i \in {\mathcal{M}}$ of MID $u$ at $t$. If $\forall j = i$, then ${c_u}(t) = 0$; otherwise ${c_u}(t) = \varepsilon$. Thus, the service migration cost of MID $u$ at $t$ can be expressed as 
\be
c_u(t)=\sum_{m=1}^{M} \frac{\varepsilon}{2} [(1-x_u^m(t-1))x_u^m(t)+(1-x_u^m(t))x_u^m(t-1)].
\ee

\section{Service Cost Minimization Problem Formulation}
We define the service cost for each MID as the weighted sum of the power consumption and service migration cost, which can be expressed as
\be
\mathcal{W}_u(t)\triangleq P_u(t)+\phi c_u(t),
\ee
where $P_u(t)\triangleq P_u^o(t)+P_u^l(t)$ is the total power consumption of MID $u$ at $t$, and $\phi \ge 0$ is the weighted coefficient of the service migration cost, which can be adjusted to address the cost of the service migration cost of MID $u$, as well as to balance the power consumption and service migration cost.

We aim to minimize the long-term average sum service cost of all the MIDs under the constraint of resource limitation and QoS requirement while guaranteeing the average queuing latency. The controller operation at $t$ is expressed as $\mathcal{O}(t)\buildrel \Delta \over=\{\mathbf{x}(t),\mathbf{p}^{tx}(t),\mathbf{f}(t)\}$. The average sum service cost minimization problem can be formulated as
\begin{subequations}
\begin{alignat}{5}
\label{P1}  
\textbf{P}_{1}{:}~ \mathop {\min}_{\mathcal{O}(t)}& \mathop {\lim }\limits_{T \to \infty } \frac{1}{T}\mathbb{E}\left[\sum_{t = 1}^T \sum_{u = 1}^U \mathcal{W}_u(t)\right]\nonumber\\	
s.t.~~& (1)-(3),\nonumber\\
& R_u(t) \ge R_{th},\\
& \mathop {\lim }\limits_{t \to \infty } \frac{1}{t}\mathbb{E}[ |\ Q_u(t)| ]=0, ~\forall u\in \mathcal{U} ,\\
&{f_u}(t) \le {f_{\max }},~\forall u \in \mathcal{U},\\
&0 \le p_u^{tx}(t) \le P_{\max }^{tx}, ~\forall u \in \mathcal{U},
\end{alignat}
\end{subequations}
where $R_u(t)\triangleq r_u^{o}(t)+r_u^{l}(t)$ is the total achievable rate of MID $u$. The constraint (9a) indicates that the total achieved rate at $t$ should be no less than the required minimum rate threshold $R_{th}$. (9b) enforces the task buffers to be mean rate stable, which also ensures that all the arrived computation tasks can be processed within a finite delay. (9c) and (9d) are the ranges of local computing frequency and the maximum allowable transmit power of each MID, respectively.
%

\section {Online Mobility-Aware Offloading and Resource Allocation Algorithm} 
\subsection{Lyapunov Optimization Framework}
To stabilize the virtual queues, we first define a quadratic Lyapunov function $L(\mathbf{Q} (t))\mathop  = \limits^\Delta  \frac{1}{2}\sum_{u=1}^U Q_u{(t)^2}$ \cite{lyapu}.
We further introduce the one-step conditional Lyapunov drift function to push the quadratic Lyapunov function towards a bounded level so that the virtual queue is stabilized. 
\be
\Delta (\mathbf{Q} (t))\mathop  = \limits^\Delta  \mathbb{E}[L(\mathbf{Q} (t + 1)) - L(\mathbf{Q} (t))|\mathbf{Q} (t)].
\ee
To incorporate queue stability, we define a Lyapunov drift-plus-penalty function \cite{lyapu} to solve the real-time problem
\be
{\Delta _V}(\mathbf{Q} (t)) = \Delta (\mathbf{Q} (t)) + V \cdot \mathbb{E}\left[\sum_{u=1}^U(P_u(t)+\phi c_u(t))|\mathbf{Q} (t)\right],
\ee
where $V $ is a control parameter greater than zero in the proposed algorithm. 
For an arbitrary feasible decision $\mathcal O(t)$ that is applicable in all the time slots, the drift-plus-penalty function ${\Delta _V}(\mathbf{Q}(t))$ must satisfy
\be
\begin{aligned}
	\Delta_V(\mathbf{Q}(t))& \le C+ \mathbb{E}\left[\sum_{u=1}^U(Q_u(t)(A_u(t)-D_{u}(t)))|\mathbf{Q} (t)\right]\\
	&+V \cdot \mathbb{E}\left[\sum_{u=1}^U(P_u(t)+\phi c_u(t))|\mathbf{Q} (t)\right],
\end{aligned} 
\ee
where $C = \frac{1}{2}\sum\limits_{u = 1}^U {({D_{u}^{\max }}^2 + {A_{u}^{max}}^2)} $, $D_{u}^{\max }$ and $A_{u}^{max}$ are the maximum achievable data and arrival workload respectively. 

The main principle of the proposed online optimization algorithm based on the Lyapunov optimization is to minimize the upper bound of $\Delta_V(\mathbf{Q}(t))$ on the right side of (12). By doing so, $\textbf{P}_1$ is converted to a series of per time slot based optimization problems. Meanwhile, the long-term average sum service cost can be minimized, and the amount of tasks waiting in the task buffers can be maintained at a low level, which effectively guarantees the average queuing latency.
The proposed algorithm is summarized in Algorithm 1, where a deterministic optimization problem $\textbf{P}_2$ needs to be solved at each time slot.
\begin{algorithm}[!t]
	\algsetup{linenosize=\small}
	\small
	\caption{ The Proposed OMORA Algorithm }
	\label{alg1}
	\begin{algorithmic}[1]
		
		\STATE At the beginning of the $t$th time slot, obtain $\{Q_u(t)\}$, $\{A_u(t)\}$.
		
		\STATE Determine $\mathbf{f}(t),\mathbf{p}^{tx}(t)$, and $\mathbf{x}(t)$ by solving 
		\be
		\begin{aligned}
			\textbf{P}_2{:}~ & \min_{\mathcal{O}(t)} \sum_{u=1}^U Q(t)[A_u(t)-D_{u}(t)]\\
			&+V\sum_{u=1}^U [P_u(t)+\phi c_u(t)]\\	
			s.t.~~& (1)-(3),(9a),(9c),(9d) \nonumber\\
		\end{aligned}
		\ee
		\STATE Update $\{Q_u(t)\}$ according to (4) and set $t=t+1$.
	\end{algorithmic}
\end{algorithm}
\subsection {Optimal Solution For $\textbf{P}_2$}
One can readily identify that $\textbf{P}_2$ is a mixed-integer programming problem involving three sets of optimization variables, namely, the local CPU-cycle frequency $\mathbf{f}(t)$, the transmit power $\mathbf{p}^{tx}(t)$, and the association indicator $\mathbf{x}(t)$. The computational complexity is prohibitively high for a brute force approach. Motivated by this, we propose to solve $\textbf{P}_2$ by optimizing these three variables alternately in an iterative way. In each iteration, the optimal local CPU-cycle frequencies and the optimal transmission power are obtained in the closed forms, and the optimal user association indicator is determined by the proposed algorithm based on semidefinite programming (SDP) relaxation. 

\textbf{Optimal Local CPU-cycle Frequencies:}
By fixing $p_u^{tx}(t)$ and $x_u^m(t)$, the optimal local CPU-cycle frequencies $\mathbf{f}(t)$ can be obtained by solving the following sub-problem $\textbf{P}_{2.1}$:
\begin{alignat}{5}\label{PC}
\textbf{P}_{2.1}{:}&\min_{ 0\le f_u(t)\le f_{max}}
V\cdot\sum_{u=1}^U \left[\kappa_{mob} f_u^3(t)\right]-Q_u(t)f_u(t)\tau/\gamma_u\nonumber\\	
&~~~~~~~s.t.~~ f_u(t)/\gamma_u\ge \max\{R_{th}-r_u^{o}(t),0\}.
\end{alignat}
Since the objective function of $\textbf{P}_{2.1}$ is a convex function, the optimal $f_u(t)$ can be derived as
\be
f_u(t)=\max\left\{(R_{th}-r_{u}^{o}(t))\gamma_u,0,\min\{\sqrt{\frac{Q_u(t)\tau}{3V\kappa_{mob} \gamma_u}}, {f_{max}}\}\right\}.
\ee
\textbf{Transmission Power Allocation:}
With a fixed associated indicator $\mathbf{x}(t)$ and local CPU-cycle frequency $\mathbf{f}(t)$, the problem $\textbf{P}_2$ is reduced to the problem $\textbf{P}_{2.2}$ given as
\begin{alignat}{5}\label{PW}
\textbf{P}_{2.2}{:}~  \min_{0\le{p}_u^{tx}(t)\le P_{max}^{tx}}&Q_u(t)(A_u(t)-\omega\tau\log_2(1+\frac{p_u^{tx}(t)H_u^m(t)}{\chi
	+ \sigma ^2}))\nonumber\\
&+V(\zeta p_u^{tx}(t)+p_r)\nonumber\\	
s.t.~~&p_u^{tx}(t)\ge (2^{\frac{R_{th}-r_u^l(t)}{\omega}}-1)\frac{\chi
	+ \sigma ^2}{H_u^m(t)}.
\end{alignat}
Since the objective function and the constraints are all convex, the solution of $p_u^{tx}(t)$ can be given as
\be
\begin{aligned}
	p_u^{tx}(t)=&\max\{(2^{\frac{(R_{th}-r_u^l(t))}{\omega}}-1)\frac{\chi
		+ \sigma ^2}{H_u^m(t)},\\
	&\min\{\frac{Q_u(t)\omega\tau\ln2}{\zeta V}-\frac{\chi
		+ \sigma ^2}{H_u^m(t)},P_{max}^{tx}\}\}.
\end{aligned}
\ee
\textbf{User Association:}
The problem $\textbf{P}_2$ can be  solved based on the given $(p_u^{tx}(t),f_u(t))$ to determine the value of the association index $x_u^m(t)$, which gives the user association result. 
%
By merging the term with respect to $x_u^m(t)$ and removing the unrelated part, problem $\textbf{P}_2$ can be transformed into
\begin{alignat}{5}
 \textbf{P}_{2.3}:\min_{x_u^m(t)}& 
\sum_{u=1}^U\sum_{m=1}^{{M}}[ \frac{1}{2}V\phi\varepsilon (1-2x_u^m(t-1))\nonumber\\
&-Q_u(t)r_u^o(t)\tau]x_u^m(t)\nonumber\\	
s.t.&~~(1)-(3).\nonumber
\end{alignat}
The problem is non-convex since the first constraint is a non-convex quadratic constraint. Here, we propose a separable Semi-Definite Program (SDP) approach to obtaining the binary association index $x_u^m(t)$ for each MID $u$ at $t$. The problem is first transformed into a convex problem based on QCQP transformation and semidefinite relaxation (SDR) to obtain the fractional solution. Then, based on the solution, the Shmoys and Tardos rounding technique is used to recover the optimal value for $x_u^m(t)$ \cite{qcqp}.

Define $\mathbf{w}_u(t)=[x_u^1(t),x_u^2(t),\cdots,x_u^M(t)]^T$ and $\mathbf{e}_m$ as the $M\times1$ standard unit vector with the $m$th entry being $1$. Let $a_u^m(t)=\frac{1}{2}V\phi\varepsilon (1-2x_u^m(t-1))-Q_u(t)r_u^o(t)\tau$, $\textbf{P}_{2.3}$ can be further transformed into an equivalent QCQP problem as follows.
\begin{subequations}
\begin{alignat}{5}
\textbf{P}_{2.3.1}&:~ \min_{\mathbf{w}_u(t)} 
\sum_{u=1}^U {\mathbf{v}^o_u}^T(t)\mathbf{w}_u(t)\nonumber\\	s.t.& \mathbf{w}_u^T(t) \text{diag}({\mathbf{e}_m})\mathbf{w}_u(t)-\mathbf{e}_m^T\mathbf{w}_u(t)=0,\forall u \in \mathcal{U},  m \in \mathcal{M},\\
& \sum_{m=1}^M \mathbf{e}_m^T\mathbf{w}_u(t)= 1,\forall u \in \mathcal{U}, t \in \mathcal{T},\\ 
&\sum\limits_{u=1}^{U}\mathbf{e}_m^T\mathbf{w}_u(t) \le{N_{max}},~\forall m \in \mathcal{M}, t \in \mathcal{T},
\end{alignat}
\end{subequations}
where $\mathbf{v}^o_u(t)=[a_u^1(t),a_u^2(t),\cdots,a_u^M(t)]^T$. The problem is still non-convex. By applying the separable SDR, the approximate solution can be obtained from the following convex problem. 

Let $\mathbf{W}_u(t)=[\mathbf{w}_u(t)^T,~1]^T\times[\mathbf{w}_u(t)^T,~1]$ and release the rank constraint $\text{rank}(\mathbf{W}_u)=1$, then the problem can be given as
\begin{subequations}
\begin{alignat}{5}
\textbf{P}_{2.3.2}:~& \min_{\mathbf{W}_u(t)} 
\sum_{u=1}^U \text{Tr}({\mathbf{V}^o_u}(t)\mathbf{W}_u(t)) \nonumber\\	
s.t.~~ &\text{Tr}(\mathbf{V}_{u,m}^x(t)\mathbf{W}_u(t))=0,\forall u \in \mathcal{U}, m \in\mathcal{M},\\
& \sum\limits_{m=1}^{M}\text{Tr}(\mathbf{V}_{u,m}^e(t)\mathbf{W}_u(t))= 1,\forall m \in\mathcal{M},\\ 
&\sum\limits_{u=1}^{U} \text{Tr}(\mathbf{V}_{u,m}^e(t)\mathbf{W}_u(t)) \le{N_{max}},~\forall m \in \mathcal{M},
\vspace{-0.1cm}
\end{alignat}
\end{subequations}
where ${\bf{V}}_u^o(t) = \left[ {\begin{array}{*{20}{c}}
	0 & {\frac{1}{2}{\bf{v}}_u^o(t)}  \\
	{\frac{1}{2}{\bf{v}}{{_u^o}^T}(t)} & 0  \\
	\end{array}} \right]$,${\bf{V}}_{u,m}^x(t) = \left[ {\begin{array}{*{20}{c}}
	\text{diag}(\mathbf{e}_m) & {-\frac{1}{2}{{\bf{e}}_m}}  \\
	{-\frac{1}{2}{\bf{e}}_m^T} & 0  \\
	\end{array}} \right]$, ${\bf{V}}_{u,m}^e(t) = \left[ {\begin{array}{*{20}{c}}
	0 & {\frac{1}{2}{{\bf{e}}_m}}  \\
	{\frac{1}{2}{\bf{e}}_m^T} & 0  \\
	\end{array}} \right]$.  \\

The problem $\textbf{P}_{2.3.2}$ is a convex problem and can be solved in a polynomial time using a standard SDP solver. Since the problem $\textbf{P}_{2.3.2}$ is a relaxation of problem $\textbf{P}_{2.3.1}$, its solution is the lower bound of the solution of the original association problem if $\text{rank}(\mathbf{W}_u^*(t))\neq 1$. To recover the  integer $x_u^m(t)$, the rounding technique \cite{qcqp} is applied as follows.
Firstly, $\mathbf{z}_u(t)=[z_u^1(t),...,z_u^M(t)]=\text{diag}(\mathbf{W}_u^*(t))$ and $z_u^m(t) \in [0,1]$ are defined as the fractional association solution of MID $u$.
After the fractional association solution of each $u$ is obtained, we then construct the weighted bipartite graph $\mathcal{G}(U, V, E)$ to establish the relationship between mobile users and MEC servers. Let $U$ denote the mobile users in the network and $V=\{ v_{m,s}:m=1,2,\cdots, M, s=1,...,J_m\}$ , where $J_m=\left\lceil {\sum\limits_{u = 1}^U {z_u^m(t)} } \right\rceil $ denotes that there are $J_m$ MIDs associated to the MEC server $m$. The nodes $\{v_{m,s}:s=1,2,...,J_m\}$ correspond to MEC server $m$. Then, the graph $G$ which includes the weighted edges between $U$ and $V$ needs to be constructed by the following  Algorithm \ref{alg_graph}.
After obtaining $\mathcal{E}$, Hungarian algorithm is utilized to find a complete max-weighted bipartite matching $M_{match}$. $M_{match}$ can be denoted as $\{(u_u,v_{m,s},e_u^{m,s}): u_u\in U,v_{m,s}\in V, e_u^{m,s} \in \mathcal{E} \}$, whose total edge weight is the maximum among all the matchings. Since this is a complete matching, each MID $u$ can find a unique matching point $v_{m,s}$.
Based on $M_{match}$, the integer user association decision can be extracted. Let $\mathbf{X}={[\mathbf{x}_1,...,\mathbf{x}_U]}^T$, where $\mathbf{x}_u=[x_{u}^{1},...,x_{u}^{M}].$ If $(u_u,v_{m,s},e_u^{m,s})$ is in $M_{match}$, set $x_u^m=1$, otherwise, $x_u^m=0$. 
\begin{algorithm}[!t]
\setlength{\abovedisplayskip}{3pt}
	\algsetup{linenosize=\small}
	\small
	\caption{\textsc{The Construction of Bipartite Graph $G$.}}
	\label{alg_graph}
	\begin{algorithmic}[1]
		
		\STATE \textbf{Set:} $\mathcal{E}\leftarrow \emptyset $.
		
		\STATE \textbf{Initialization:}Sort the $z_u^m(t)$ in descent order for each $m$ as $x^{'m}_1(t)\ge x^{'m}_2(t)\ge \cdots \ge x^{'m}_U(t)$ for each $m$.
		\IF {$J_m\le 1$}
		\FOR {each $x_u^{'m} \ge 0$}
		
		\STATE Add the edge $(u_u,v_{m,1})$ into $\mathcal{E}$ and set $e_{u}^{m,1}=x_u^{'m}$.	
		\ENDFOR
		\ELSE
		\FOR {$s=1,2,...,J_m$}
		\STATE find the minimum index $u_s$ such that $\sum_{u=1}^{u_s}x_u^{'m}\ge s$. 
		\IF {$u=u_{s-1}+1,...,u_s-1$, and $x_u^{'m}\ge 0$}
		\STATE Add edge $(u_u,v_{m,s})$ into $\mathcal{E}$ with weight $e_{u}^{m,s}=x_u^{'m}$.
		\ELSIF{$u=u_s$}
		\STATE Add edge $(u_u,v_{m,s})$ into $\mathcal{E}$ with weight $e_{u}^{m,s}=1-\sum_{u=u_{s-1}+1}^{u_s-1}x_u^{'m}$. 
		\STATE Add edge  $(u_u,v_{m,s+1})$ into $\mathcal{E}$ with weight $e_{u}^{m,s+1}=\sum_{u=1}^{u_s}x_u^{'m}-s$.
		\ENDIF
		\ENDFOR
		\ENDIF
		\STATE \textbf{Output:} $\mathcal{E}$.
	\end{algorithmic}
\end{algorithm}

\section{Simulation Results}
\label{Simulation}
In this section, simulation results are provided to evaluate the proposed algorithm. The simulation settings are  based on the work in \cite{singlemao},\cite{qunwang}. There are $3$ MEC servers and $10$ MIDs randomly deployed in a $100\times 100~m^2$ area. The MID trajectory is generated by the random walk model. The arrival workload $A_u(t)$ is uniformly distributed within $[0.95,1.5]\times 10^6$ bits. 
The channel power is exponentially distributed with the mean of
$g_0\cdot (d/d_0)^{-4}$, where the reference distance $d_0=1$ and $g_0 =-40$ dB. $\kappa_{mob} = 10^{-28}$, $\omega= 1$ MHz, $\sigma^2 = 10^{-13}$ W, $\chi=10^{-10}$ W,
$P_{max} = 1$ W, $f_{max} = 2.15$ GHz, $\gamma_u=737.5$ cycles/bit, the amplifier coefficient $\zeta=1$, $\varepsilon=10^{-1}$, $\phi=0.1$.  \\

We consider two cases as benchmarks to evaluate the proposed algorithm. The first benchmark, marked as "NL",  has no local computation but has a dynamic user association policy. The second benchmark, marked as "NM", has  no service migration by keeping initial association decision unchanged.
\begin{figure}[h]
	\centering
	\includegraphics[width=3.8in]{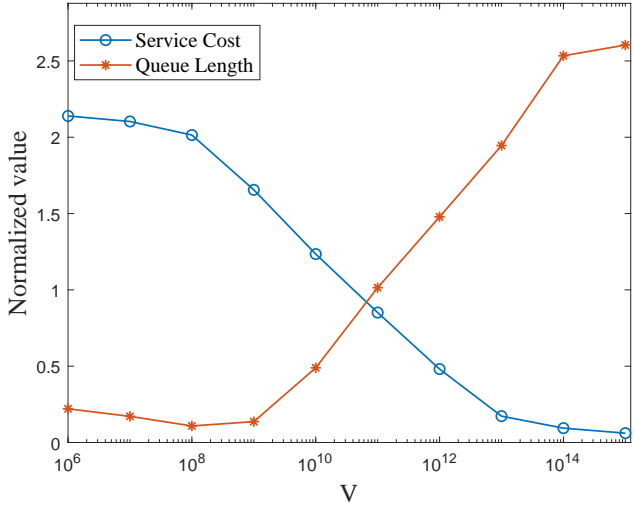}	\label{qvc}
	\caption{Service cost/queue length v.s. control parameter $V$.}
\end{figure}
\begin{figure}[h]
	\centering
	\includegraphics[width=3.8in]{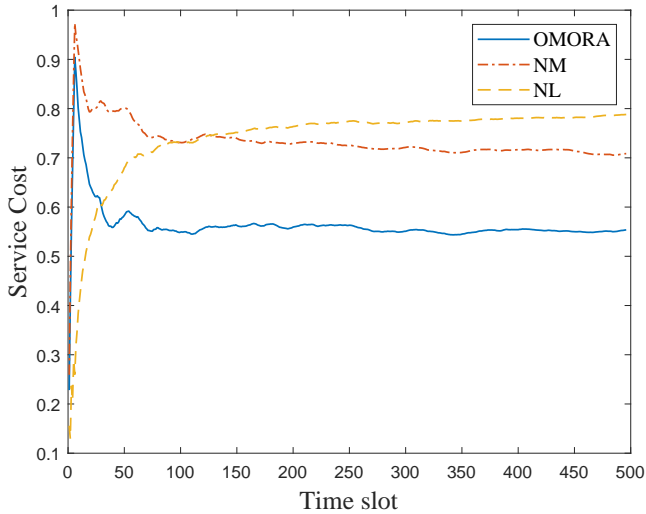}	\label{slotsy}
	\caption{Service cost.}
\end{figure}
\begin{figure}[h]
	\centering
	\includegraphics[width=3.8in]{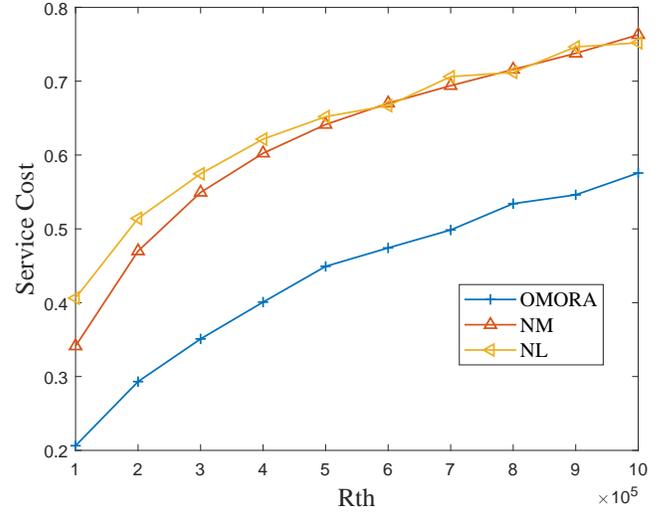}	\label{rth}
	\caption{Service cost v.s. minimum required rate $R_{th}$.}
\end{figure}

A comparison of the achievable service cost/task queue length versus the control parameter $V$ is presented in Fig. 2. 
The service cost and task queue length are first maintained at a stable level when $V$ is small. 
With the increase of $V$, the system gives more incentive to minimize the service cost than to lower down  the queue length. Thus, the service cost decreases and the queue length increases. The best trade-off to maintain a lower service cost, as well as the lower queue length occurs around $V=10^{10}$. Therefore, in the following simulation, the control parameter $V$ is set as $10^{10}$.

A comparison of the average service cost versus time slot is presented in Fig. 3. As shown in the figure, the proposed method can achieve the lowest service cost compared with the other two methods. This is because with the assistance of local processing and user association, the system can save more power through local computing and receive a better service from MEC.

The service cost versus the minimum computation rate requirement $R_{th}$ is presented in Fig. 4. The proposed method can achieve the lowest service cost. With the increase of $R_{th}$, the computation rate constraint forces each method to increase its powers, which causes a higher service cost. It is worth noting that the gaps between those three methods keep increasing with the increase of $R_{th}$. The reason is that when the required rate is at a low level, all the methods  consume a lower power to meet the service requirement. However, when the rate keeps increasing, the system needs to allocate more power and to choose the best method to achieve a lower service cost. Therefore, the proposed method is more adaptable and can achieve a better performance.
 \begin{figure}[h]
	\centering
	\includegraphics[width=3.8in]{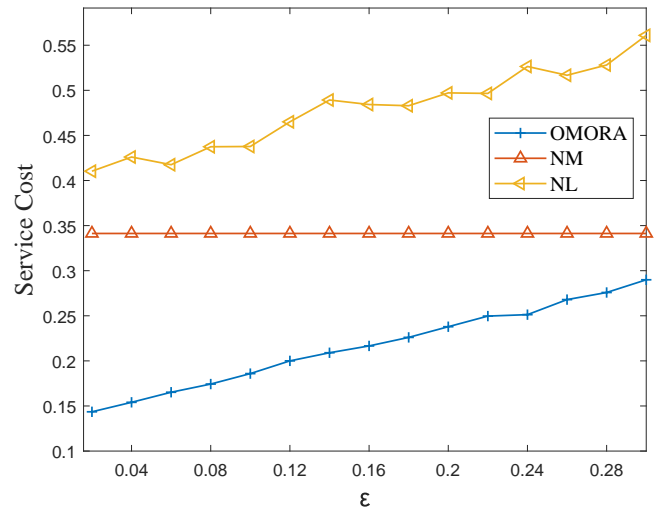}	\label{eth}
	\caption{Service cost v.s. migration cost $\varepsilon$.}
\end{figure}

Fig.5 illustrates the average system service cost versus migration cost $\varepsilon$. It can be seen that the average service cost of all the methods except ‘NM’ increases with $\varepsilon$, which is in accordance with our intuition. The service cost of the proposed scheme keeps increasing and finally approaches the service cost of “NM” when the migration cost is at a large value. This observation confirms that our proposed scheme can achieve a better trade-off between service migration cost and energy consumption. \\

\section{Conclusions}
In this paper, we investigated task offloading and resource allocation in an MEC-enabled IoT network. The average service cost minimization problem with QoS constraint and the task queue stability constraint was formulated as a stochastic optimization problem. A mobility-aware task offloading and resource allocation algorithm based on Lyapunov optimization and SDP was developed, which jointly optimizes the transmit power, the CPU-cycle frequencies, and the user association vector of IoT MIDs. It was shown that the proposed algorithm outperforms other benchmarks and is capable of balancing the service cost and the delay performance in a MEC-enabled IoT network with mobility consideration. \\

\end{document}